\documentclass{natureprintstyle}
\usepackage{graphicx}
\usepackage{amssymb,hyperref}

\usepackage{multibib}
\newcites{meth}{\mbox{ }}

\bibliographystylemeth{naturemag-swj}

\newcommand\arcsec{\mbox{$^{\prime\prime}$}}%
\title{Flows of X-ray gas reveal the disruption of a star by a massive black hole}

\author{Jon M. Miller$^{1}$, 
Jelle S. Kaastra$^{2,3,4}$, 
M. Coleman Miller$^{5}$,
Mark T. Reynolds$^{1}$,
Gregory Brown$^{6}$, 
S. Bradley Cenko$^{7,8}$,
Jeremy J. Drake$^{9}$,
Suvi Gezari$^{5}$,
James Guillochon$^{10}$,
Kayhan Gultekin$^{1}$,
Jimmy Irwin$^{11}$,
Andrew Levan$^{6}$,
Dipankar Maitra$^{12}$,
W. Peter Maksym$^{11}$,
Richard Mushotzky$^{5}$,
Paul O'Brien$^{13}$,
Frits Paerels$^{14}$,
Jelle de Plaa$^{2}$,
Enrico Ramirez-Ruiz$^{15}$,
Tod Strohmayer$^{7}$,
Nial Tanvir$^{13}$
}

\begin{document}

\maketitle

\begin{affiliations}
\item Department of Astronomy, The University of Michigan, 1085 South
  University Avenue, Ann Arbor, Michigan, 48103, USA.
\item SRON Netherlands Institute for Space Research, Sorbonnelaan 2,
  3584 CA Utrecht, The Netherlands.
\item Department of Physics and Astronomy, Universiteit Utrecht, PO
  BOX 80000, 3508 TA Utrecht, The Netherlands.
\item Leiden Observatory, Leiden University, PO BOX 9513, 2300 RA
  Leiden, The Netherlands.
\item Department of Astronomy, The University of Maryland, College
  Park, Maryland, 20742, USA.
\item Department of Physics, University of Warwick, Coventry, CV4 7AL, UK.
\item Joint Space-Science Institute, University of Maryland, College Park, MD, 02742, USA.
\item Astrophysics Science Division, NASA Goddard Space Flight Center,
  MC 661, Greenbelt, Maryland, 20771, USA.
\item Smithsonian Astrophysical Observatory, 60 Garden Street,
  Cambridge, Massachusetts, 02138, USA.
\item The Institute for Theory and Computation, Harvard-Smithsonian
  Center for Astrophysics, 60 Garden Street, Cambridge, Massachusetts,
  02138, USA.
\item Department of Physics and Astronomy, University of Alabama, BOX
  870324, Tuscaloosa, Alabama, 35487, USA.
\item Department of Physics and Astronomy, Wheaton College, Norton,
  Massachusetts, 02766, USA.
\item Department of Physics and Astronomy, University of Leicester,
  University Road, Leicester, LE1 7RH, UK.
\item Columbia Astrophysics Laboratory and Department of Astronomy,
  Columbia University, 550 West 120th Street, New York, New York,
  10027, USA.
\item Department of Astronomy and Astrophysics, University of
  California, Santa Cruz, California, 95064, USA.
\end{affiliations}

\begin{abstract}
Tidal forces close to massive black holes can violently disrupt stars
that make a close approach.  These extreme events are discovered via
bright X-ray\cite{Bade96,Komossa99,Esquej07,Cappelluti09} and
optical/UV\cite{Gezari08,Gezari12} flares in galactic centers.  Prior
studies based on modeling decaying flux trends have been able to
estimate broad properties, such as the mass accretion
rate\cite{Gezari12,Guillochon14}.  Here we report the detection of
flows of highly ionized X-ray gas in high-resolution X-ray spectra of
a nearby tidal disruption event.  Variability within the
absorption-dominated spectra indicates that the gas is relatively
close to the black hole.  Narrow line widths indicate that the gas
does not stretch over a large range of radii, giving a low volume
filling factor.  Modest outflow speeds of $few \times 10^{2}~ {\rm
  km}~ {\rm s}^{-1}$ are observed, significantly below the escape
speed from the radius set by variability.  The gas flow is consistent
with a rotating wind from the inner, super-Eddington region of a
nascent accretion disk, or with a filament of disrupted stellar gas
near to the apocenter of an elliptical orbit.  Flows of this sort are
predicted by fundamental analytical theory\cite{Rees88} and more
recent numerical
simulations\cite{Strubbe09,Lodato09,Lodato11,Strubbe11,Guillochon14,Shiokawa15,Cole15}.
\end{abstract}

ASASSN-14li was discovered in images obtained on November 22, 2014
(MJD 56983), at a visual magnitude of V=16.5\cite{Jose14} by the
All-Sky Automated Survey for Supernovae (ASAS-SN).  Follow-up
observations found the transient source to coincide with the center of
the galaxy PGC 043234 (originally Zw VIII 211), to within 0.04 arc
seconds\cite{Jose14}.  This galaxy lies at a red-shift of $z =
0.0206$, or a luminosity distance of 90.3 Mpc (for ${\rm H}_0 =
73~{\rm km}~ {\rm s}^{-1}$, $\Omega_{\rm matter} = 0.27$,
$\Omega_{\Lambda} = 0.73$), making ASASSN-14li the closest disruption
event discovered in over 10 years.  The discovery magnitudes indicated
a substantial flux increase over prior, archival optical images of
this galaxy.  Follow-up observations with the \emph{Swift} X-ray
Telescope\cite{Gehrels04,Burrows05} (XRT) established a new X-ray
source at this location\cite{Jose14}.

Archival X-ray studies rule out the possibility that PGC 043234
harbours a standard active galactic nucleus that could produce bright
flaring.  PGC 043234 is not detected in the ROSAT All-Sky
Survey\cite{Voges99}.  Utilizing the online interface to the data, the
background count rate for sources detected in the vicinity is $0.002~
{\rm counts}~ {\rm s}^{-1}~ {\rm arcmin}^{-2}$.  With standard
assumptions (see \emph{Methods}), this rate corresponds to a
luminosity of $L \simeq 4.8\times 10^{40}~ {\rm erg}~ {\rm s}^{-1}$,
which is orders of magnitude below a standard active nucleus.

Theory predicts that early tidal disruption event (TDE) evolution
should be dominated by a bright, super-Eddington accretion phase, and
be followed by a charactecteristic $t^{-5/3}$ decline as disrupted
material interacts and accretes\cite{Rees88,Phinney89}.  Detections of
winds integral to super-Eddington accretion have not been reported
previously, but $t^{-5/3}$ flux decay trends in the UV (where disk
emission from active nuclei typically peaks) are now a standard
signature of TDEs in the literature\cite{Gezari08,Gezari12}.  Figure 1
shows the flux decay of ASASSN-14li, as observed by \emph{Swift}.  A
fit to the UVM2 data assuming an index of $\alpha = -5/3$ gives a
disruption date of $t_0 \simeq 56948\pm 3$ (MJD).  The
V-band light is consistent with a shallower $t^{-5/12}$ decay; this
can indicate direct thermal emission from the disk, or reprocessed
emission\cite{Lodato11,Guillochon14} (see \emph{Methods}).

We triggered approved \emph{XMM-Newton} programs to study ASASSN-14li
soon after its discovery.  Although \emph{XMM-Newton} carries several
instruments, the spectra from the two RGS units are the focus of this
analysis.  We were also granted a Director's Discretionary Time
observation with \emph{Chandra}, using its Low Energy Transmission
Grating spectrometer (LETG), paired with its High Resolution Camera
for spectroscopy (HRC-S).

The 18--35\AA~ X-ray spectra of ASASSN-14li are clearly thermal in
origin, so we modeled the continuum with a single blackbody, modified
by interstellar absorption in PGC 043234 and the Milky Way, and
absorption from blue-shifted, ionized gas local to the TDE.  The
self-consisent photoionization code ``pion''\cite{Kaastra96} was used
to model the complex absorption spectra (see Table 1, and
\emph{Methods}).

Assuming that the highest bolometric luminosity derived in fits to the
high-resolution spectra ($L = 3.2\pm 0.1 \times 10^{44}~ {\rm erg}~
{\rm s}^{-1}$) corresponds to the Eddington limit, a black hole mass
of $2.5\times 10^{6}~ M_{\odot}$ is inferred.  The blackbody emission
measured in fits to the time-averaged \emph{XMM-Newton} spectrum gives
an emitting area of $3.7\times 10^{25}~ {\rm cm}^{2}$; implying $r =
1.7\times 10^{12}~ {\rm cm}$ for a spherical geometry.  This is
consistent with the innermost stable circular orbit (ISCO) around an
$M \simeq 1.9\times 10^{6}~ M_{\odot}$ black hole.  Modeling of the
{\it Swift} light curves (see Figure 1) using a self-consistent
treatment of direct and reprocessed light from an elliptical accretion
disk\cite{Guillochon14} gives a mass in the range of $M \simeq 0.4-1.2
\times 10^{6}~ M_{\odot}$ (please see the \emph {Methods}).  In
concert, the thermal spectrum, implied radii, and the run of emission
from X-rays to optical bands unambiguously signal the presence of an
accretion disk in ASASSN-14li.

Figure 2 shows the best-fit model for the spectra obtained in the long
stare with the \emph{XMM-Newton}/RGS (see Table 1, and \emph{Methods}).  An F-test
finds that photoionized X-ray absorption is required in fits to these
spectra at more than the 27$\sigma$ level of confidence, relative to a
spectral model with no such absorption.  The model captures the
majority of the strong absorption lines, giving $\chi^{2} = 870.5$ for
563 degrees of freedom (see Table 1).  The strongest lines in the
spectrum coincide with ionized charge states of N, O, S, Ar, and Ca.
Only solar abundances are required to describe the spectra.  The
\emph{Chandra} spectrum independently confirms these results in broad
terms, and requires absorption at more than the 6$\sigma$ level of
confidence.

A hard lower limit on the radius of the absorbing gas is set by the
the blackbody continuum.  The best radius estimate likely comes
from variability time scales within the \emph{XMM-Newton} long stare.
Analysis of specific time segments within the long stare, as well as
flux-selected segments, reveals that the absorption varies (see Table 1, and
\emph{Methods}).  This sets a relevant limit of $r \leq c \delta t$,
or $r \leq 3\times 10^{15}~ {\rm cm}$.  While the column density and
ionization do not vary significantly, the blue-shift of the gas does.
During the initial third of the observation, the blue-shift is larger,
$v_{\rm shift} = -360\pm 50~ {\rm km}~ {\rm s}^{-1}$, but falls to
$v_{\rm shift} = -130^{-50}_{+70}~ {\rm km}~ {\rm s}^{-1}$ in the
final two-thirds.  Shorter monitoring observations with
\emph{XMM-Newton} reveal evolution of the absorbing gas, including
changes in ionization and column density, before and after the long
stare (see Table 1, and \emph{Methods}).

Fundamental theoretical treatments of TDEs predict an initial
near-Eddington or super-Eddington phase\cite{Rees88}; this is
confirmed in more recent theoretical
studies\cite{Loeb97,Strubbe09,Strubbe11}.  The high-resolution X-ray
spectra were obtained within the predicted time frame for
super-Eddington accretion, for our estimates of the black hole
mass\cite{Piran15}.  Although the ionization parameter of the observed
gas is high, the ionizing photon distribution peaks at a low energy,
and the wind could be driven by radiation force.  Such flows are
naturally clumpy, and may be similar to the photospheres of
novae\cite{Shaviv01}.  Given the strong evidence of an accretion disk
in our observations of ASASSN-14li, the X-ray outflow is best
associated with a wind from the inner regions of a nascent,
super-Eddington accretion disk.  The local escape speed at an
absorption radius of $r \simeq 10^{4}~ {\rm GM/c^{2}}$ (appropriate
for $M \simeq 10^{6}~ M_{\odot}$) exceeds the observed outflow
line-of-signt speed of the gas, but Keplerian rotation is not encoded
in absorption, and projection effects are also important.  The small
width of the absorption lines relative to the escape velocity may also
indicate a low filling factor, consistent with a clumpy outflow or
shell.

The existing observations show a general trend toward higher outflow
speeds with time.  Corresponding changes in ionization and column
density are more modest, and not clearly linked to outflow
speed\cite{Ramirez11}.  However, some recent work has predicted higher
outflow speeds in an initial super-Eddington disk regime, and lower
outflow speeds in a subsequent thin disk
regime\cite{Strubbe09,Strubbe11}.  An observation
in an earlier, more highly super-Eddington phase might have observed
broader lines and higher outflow speeds; future observations of new
TDEs can test this.

Figure 3 shows the time evolution of the blackbody temperature
measured in \emph{Swift}/XRT monitoring observations.  The temperature
is remarkably constant, especially in contrast to the optical/UV decline
shown in Figure 1.  Observations of steady blackbody temperatures
despite decaying multi-wavelength light curves in some
TDEs\cite{Gezari12,Holoien14} has recently been explained through
winds\cite{Cole15}.  Evidence of winds in our data supports this
picture.

The low gas velocities may also be consistent with disrupted stellar
gas on an elliptical orbit in a nascent disk, near apocenter.  This picture naturally gives a
low filling factor, resulting in a small total mass in absorbing gas
(see \emph{Methods}).  Recent numerical simulations predict that a
fraction of the disrupted material in a TDE will circularize
slowly\cite{Shiokawa15}, and that flows will be
filamentary\cite{Guillochon15}, while stellar gas that is more tightly
bound can form an inner, Eddington-limited or super-Eddington disk
more quickly.  

The highly ionized, blue-shifted gas discovered in our high-resolution
X-ray spectra of ASASSN-14li confirms both fundamental and very recent
theoretical predictions for the structure and evolution of tidal
disruption events.  The field can now proceed to pair high-resolution
X-ray spectroscopy with an ever-increasing number of TDE detections to
test models of accretion disk formation and evolution, and to
explore strong-field gravitation around massive black holes\cite{Stone12}.

\bibliography{ms3}

\begin{addendum}

\item[Acknowledgments]
We thank \emph{Chandra} Director Belinda Wilkes and the \emph{Chandra}
team for accepting our request for Director's Discretionary Time,
\emph{XMM-Newton} Director Norbert Schartel and the \emph{XMM-Newton}
team for executing our approved target-of-opportunity program, and
\emph{Swift} Director Neil Gehrels and the \emph{Swift} team for
monitoring this important source.  J.M.M. is supported by NASA
funding, through \emph{Chandra} and \emph{XMM-Newton} guest observer
programs.  SRON is supported by NWO, the Netherlands Organization for
Scientific Research.  JJD was supported by NASA Contract NAS8-03060 to
the Chandra X-ray Center.  W.P.M. is grateful for support by the
University of Alabama Research Stimulation Program.

\item[Author Contributions] 
J.M.M. led the \emph{Chandra} and \emph{XMM-Newton} data reduction and
analysis, with contributions from J.S.K., J.J.D., and J.P.  M.R. led
the \emph{Swift} data reduction and analysis, with help from B.C.,
S.G, and R.M.  M.C.M., E.R.-R., and J.G. provided theoretical
insights.  G.B., K.G., J.I., A.L., D.M., W.P.M., P.O., D.P., F.P., T.S.,
and N.T. contributed to the discussion and interpretation.

\item[Competing Interests] 
The authors declare that they have no competing financial interests.

\item[Correspondence] 
Correspondence and requests for materials should be addressed to
J.M.M. (jonmm@umich.edu).
\end{addendum}

\clearpage

\begin{table*}
\centering
\caption{\textbf{Modeling of the high-resolution X-ray spectra reveals
    ionized flows of gas.} Each spectrum was fit with a simple
  blackbody continuum, modified by photoionized absorption via the
  ``pion'' model, and interstellar absorption in the host galaxy PGC
  043234 and the Milky Way.  The fits were made using
  ``SPEX''\cite{Kaastra05}, minimizing a $\chi^{2}$ statistic.  In all
  cases, $1\sigma$ errors are quoted.  Where a parameter is quoted
  with an asterisk, the listed parameter was not varied.  X-ray fluxes
  and luminosities listed with the suffix ``b'' for \emph{broad} were
  extrapolated from the fitting band to the 1.24--124\AA~ band; those
  with the suffix ``f'' represent values for the 18--35\AA~
  \emph{fitting} band.  Interstellar column densities are separately
  measured for the Milky Way ($N_{\rm H, MW}$) at zero red-shift, and
  the host galaxy PGC 043234 ($N_{\rm H, HG}$) at red-shift of
  $z=0.0206$.  These parameters were measured in the \emph{XMM-Newton}
  long stare and then fixed in fits to other spectra.  Variable
  parameters in the photoionization model are listed together; the
  negative $v_{\rm shift}$ values indicate a blue shift relative to
  the host galaxy.}
\label{tab:results}
\bigskip
\scriptsize
\sffamily
\begin{tabular}{lllllll}
\hline
\hline
Mission  & \emph{XMM-Newton} & \emph{XMM-Newton} & \emph{XMM-Newton} & \emph{XMM-Newton} & \emph{Chandra}  & \emph{XMM-Newton} \\

ObsId    & 0694651201 & 0722480201 & 0722480201 & 0722480201 & 17566, 17567 & 0694651401 \\

comment  & monitoring & long stare & stare (low) & stare (high) & -- & monitoring \\

Start (MJD) & 56997.98 & 56999.54 & 56999.94 & 57000.0 & 56999.97, 57002.98 & 57023.52 \\

Duration (ks) & 22 & 94 & 36 & 58 & 35, 45 & 23.6 \\

\hline 

${\rm F}_{\rm X, b}$ ($10^{-11}~ {\rm erg}~ {\rm cm}^{-2}~ {\rm s}^{-1}~$) & $2.7\pm 0.7$ & $3.2\pm 0.4$ & $3.4\pm 0.3$ & $3.4\pm 0.2$ & $2.5^{+0.2}_{-0.3}$ & $2.68\pm 0.08$ \\

${\rm L}_{\rm X, b}$ ($10^{44}~ {\rm erg}~ {\rm s}^{-1}$) & $2.9\pm 0.7$ & $2.2\pm 0.3$ & $2.2\pm 0.2$ & $2.0\pm 0.1$ & $1.7_{-0.2}^{+0.1}$ & $3.2\pm 0.1$ \\

\hline

${\rm F}_{\rm X, f}$ ($10^{-11}~ {\rm erg}~ {\rm cm}^{-2}~ {\rm s}^{-1}~$) & $1.2\pm 0.3$ & $1.2\pm 0.2$ & $1.07\pm 0.08$ & $1.24\pm 0.08$ & $1.0^{+0.1}_{-0.2}$ & $1.19\pm 0.04$ \\

${\rm L}_{\rm X, f}$ ($10^{44}~ {\rm erg}~ {\rm s}^{-1}$) & $0.25\pm 0.06$ & $0.21\pm 0.03$ & $0.19\pm 0.01$ & $0.21\pm 0.01$ & $0.17_{-0.02}^{+0.01}$ & $0.27\pm 0.01$ \\

\hline

${N}_{\rm H, MW} (10^{20}~ {\rm cm}^{-2})$ & 2.6* & $2.6\pm 0.6$ & 2.6* & 2.6* & 2.6* & 2.6*  \\

${N}_{\rm H, HG} (10^{20}~ {\rm cm}^{-2})$ & 1.4* & $1.4\pm 0.5$ & 1.4* & 1.4* & 1.4* & 1.4*  \\

\hline

${N}_{\rm H, TDE} (10^{22}~ {\rm cm}^{-2})$ & $0.7\pm 0.2$ & $1.3^{+0.9}_{-0.4}$ & $0.1^{+0.3}_{-0.2}$ & $0.9^{2}_{-0.3}$ & $0.5^{+0.4}_{-0.1}$ & $0.5\pm 0.1$ \\

log($\xi$) (erg~cm~s$^{-1}$) & $3.6\pm 0.1$ & $4.1\pm 0.2$ & $4.1\pm 0.1$ & $3.9^{+0.3}_{-0.1}$ & $3.9^{+0.1}_{-0.2}$ & $3.7\pm 0.1$ \\

$v_{\rm rms}$ (km $s^{-1}$) & $130\pm 30$ & $110^{+30}_{-20}$ & $60^{+60}_{-50}$ & $120\pm 20$ & $120^{+40}_{-30}$ & $230^{+60}_{-50}$ \\

$v_{\rm shift}$ (km $s^{-1}$) & $-180\pm 60$ & $-210\pm 40$ & $-360\pm 50$ & $-130^{-50}_{+70}$ & $-500_{-70}^{+60}$ & $-490\pm 70$ \\

\hline

kT (eV) & $50.0\pm 0.09$ & $51.4\pm 0.1$ & $50.0\pm 0.4$ & $52.6\pm 0.4$ & $52.6\pm 0.3$ & $49.7\pm 0.9$ \\

Norm ($10^{25}~ {\rm cm}^{2}$) & $5.7\pm 1.4$ & $3.7\pm 0.5$ & $4.0\pm 0.3$ & $3.0\pm 0.2$ & $2.5^{+0.1}_{-0.2}$ & $6.1\pm 0.2$ \\

\hline

$\chi^{2}/\nu$ & 704.8/567 & 870.5/563 & 687.8/564 & 726.8/565 & 266.5/178 & 626.5/566 \\
\hline
\hline
\end{tabular}
\end{table*}

\clearpage

\begin{figure*}
\begin{center}
\includegraphics[width=5.8in]{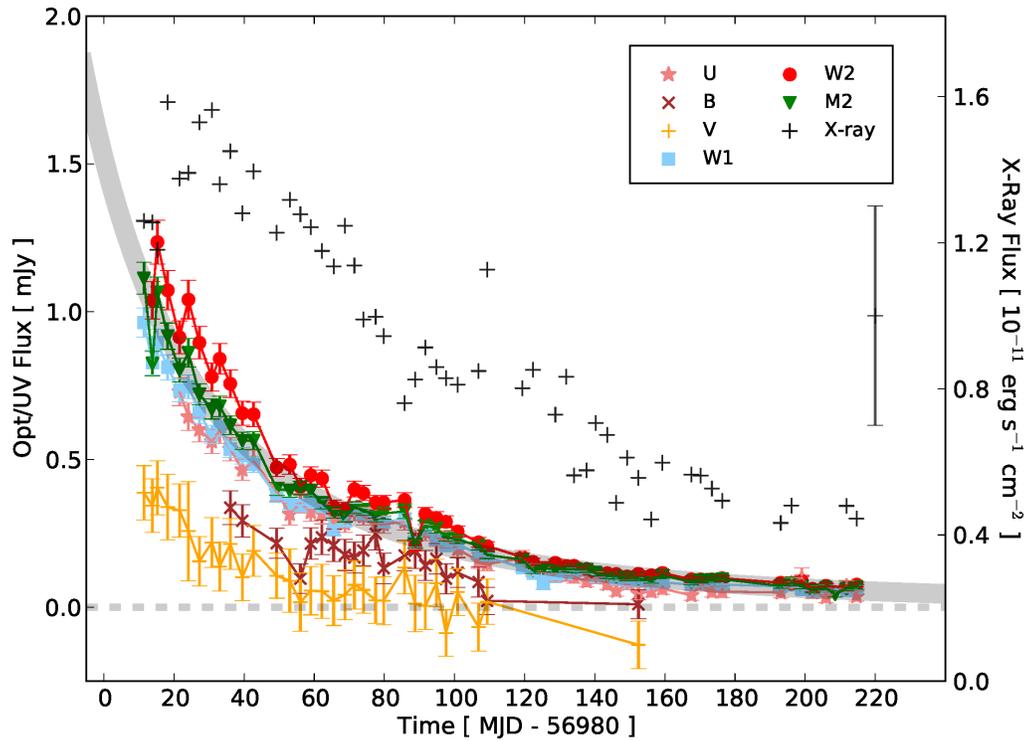}
\end{center}
\vskip -0.2in
\caption{\textbf{The multi-wavelength light curves of ASASSN-14li clealy signal a tidal disruption event.}  The light curves are based on monitoring
  observations with the \emph{Swift} satellite.  The errors shown are
  the $1\sigma$ confidence limits on the flux in each band.
  Contributions from the host galaxy have been subtracted (see \emph{
    Methods}).  The UVM2 filter samples the UV light especially well.
  The gray band depicts the $t^{-5/3}$ flux decay predicted by
  fundamental theory\cite{Rees88,Phinney89}.  The X-ray flux points
  carry relatively large errors; a representative error is shown.
  Fits to the decay curve are described in the main text and in the
  {\emph Methods}.
  \label{fig:lcurve}}
\end{figure*}

\clearpage

\begin{figure*}
\begin{center}
\includegraphics[width=6.5in]{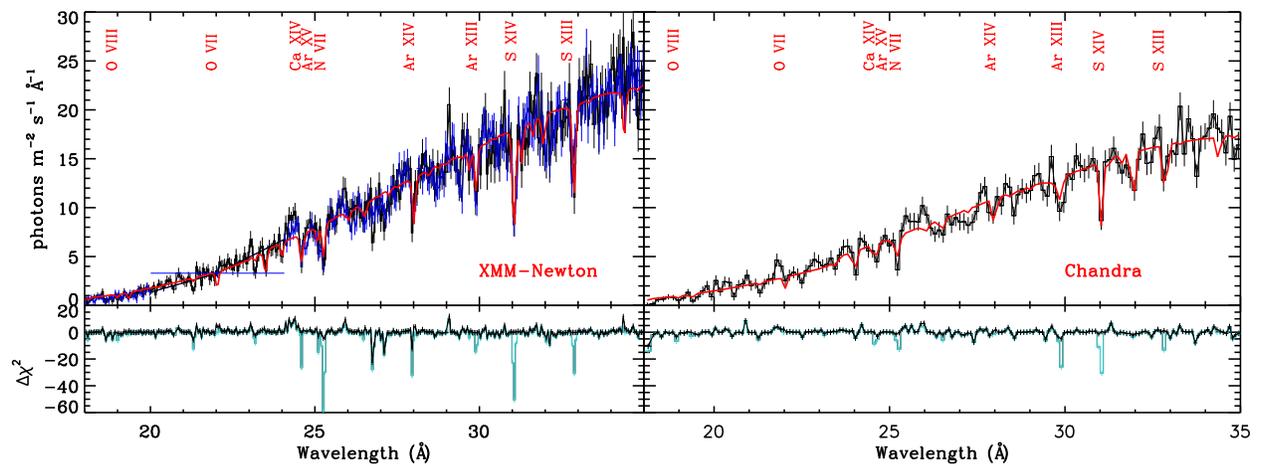}
\end{center}
\vskip -0.2in
\caption{\textbf{The high-resolution X-ray spectra of ASASSN-14li
    reveal blue-shifted absorption lines.}  Spectra from the long
  stare with {\it XMM-Newton} and the combined {\it Chandra} spectrum
  are shown.  {\it XMM-Newton} spectra from the RGS1 and RGS2 units
  are shown in black and blue, respectively; the RGS2 unit is missing
  a detector in the 20--24\AA~ band.  The best-fit photoionized
  absorption model for the outflowing gas detected in each spectrum is
  shown in red (see {\it Methods}), and selected strong lines are
  indicated.  Below each spectrum, the goodness-of-fit statistic
  ($\Delta \chi^{2}$) is shown before (cyan) and after (black)
  modeling the absorbing gas.\label{fig:big}}
\end{figure*}

\clearpage

\begin{figure*}
\begin{center}
\includegraphics[width=3.0in,angle=-90]{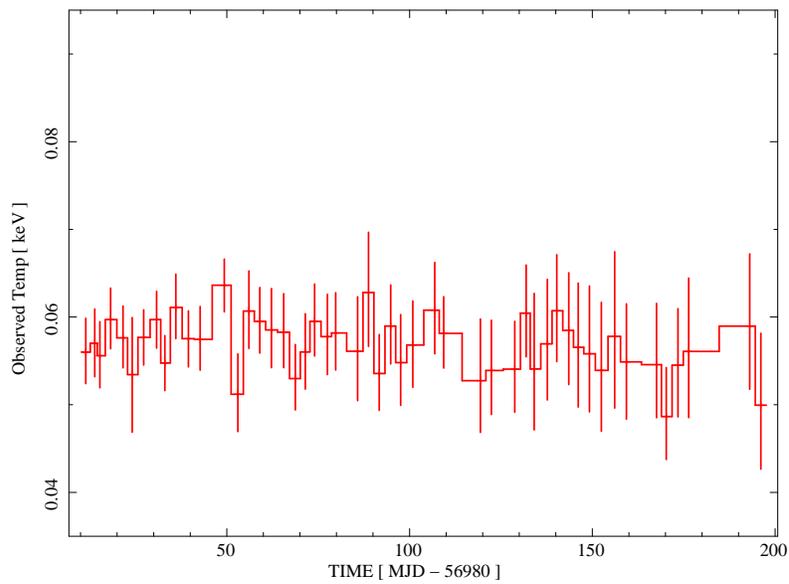}
\end{center}
\vskip -0.2in
\caption{\textbf{The temperatue of blackbody continuum emission from
    ASASSN-14li is steady over time.}  The temperature measured in
  simple blackbody fits to \emph{Swift}/XRT monitoring observations is
  plotted versus time.  Errors are 1$\sigma$ confidence intervals.
  The temperature is remarkably steady, contrasting strongly with the
  declining fluxes shown in Figure 1.  Recent theory suggests that
  winds may serve to maintain steady temperatures in some
  TDEs\cite{Cole15}.
  \label{fig:ktbb}}
\end{figure*}

\clearpage

\section*{Methods}

\mbox{ }

\vskip 0.05in 
\noindent 


\noindent \textbf{Estimates of prior black hole luminosity} 

Utilizing the ROSAT All-Sky Survey\citemeth{Voges99}, the region
around the host galaxy, PGC 043234, was searched for
point sources.  No sources were found.  Points in the vicinity of the
host galaxy were examined to derive a background count rate of 0.002
counts ${\rm s}^{-1}$.  Assuming the Milky Way column density along
this line of sight, and taking a typical Seyfert X-ray spectral index
of $\Gamma = 1.7$, this count rate translates into $L \simeq 4.8\times
10^{40}~ {\rm erg}~ {\rm s}^{-1}$.  This limit is orders of magnitude
below a Seyfert or quasar luminosity.\\ 

\noindent \textbf{Optical/UV monitoring observations and data
  reduction}

\emph{Swift}\citemeth{Gehrels04} monitors transient and variable
sources via co-aligned X-ray (XRT: 0.3 - 10 keV) and UV-Optical (UVOT:
170-650 nm) telescopes.  High-cadence monitoring of ASASSN-14li with
UVOT has continued in six bands: V, B, U, UVW1, UVM2,
and UVW2 ($\lambda_c \sim 550, 440, 350, 260, 220, 190~{\rm nm}$).

All observations were processed using the latest HEASOFT suite and
calibrations. Individual optical/UV exposures were astrometrically
corrected and sub-exposures in each filter were summed.  Source fluxes
were then extracted from an aperture of 3\arcsec radius, and
background fluxes were extracted from a source-free region to the east
of ASASSN-14li due to the presence of a (blue) star lying 10 arc
seconds to the South, using \textsc{uvotmaghist}.

To estimate the host contamination, we have measured the host flux in
3\arcsec aperture (matched to the aperture used for the UVOT
photometry) in pre-outburst Sloan Digital Sky Survey (SDSS;
\citemeth{aaa14}), 2 Micron All-Sky Survey (2MASS; \citemeth{scs06}),
and \textit{GALEX} \citemeth{mfs05} images.  Extra caution was used to
deblend the \textit{GALEX} data, where large PSF resulted in
contamination from the star $\sim 10$\arcsec to the South.  We
estimated the uncertainty in each host flux by varying the inclusion
aperture from 2\arcsec to 4\arcsec.

We then fit the host photometry to synthetic galaxy
templates using the Fitting and Assessment of Synthetic Templates
(FAST; \citemeth{kvl09}) code.  We employed stellar templates
from the \citemeth{bc03} catalog, and allowed the star formation
history, extinction law, and initial mass function to vary 
over the full range of parameters allowed by the software.  All
best fit models had stellar masses $\approx 10^{9.2}$\,$M_{\odot}$,
low ongoing star formation rates 
(SFR $\lesssim 10^{-1.5}$\,$M_{\odot}$\,yr$^{-1}$), and modest
line-of-sight extinction ($A_{V} \lesssim 0.4$\,mag).  

We took the resulting galaxy template spectra and integrated 
these over each UVOT filter bandpass to estimate the host
count rate.  For the uncertainty in this value, we adopt 
either the root-mean-square spread of the resulting galaxy
template models, or 10\% of the inferred count rate, whichever
value was larger.  We then subtracted these values from
our measured (coincidence-loss corrected) photometry of the host 
plus transient, to isolate the component due to TDE.  For
reference, our inferred count rates for each UVOT filter are:
$V = 5.7 \pm 0.6$\,s$^{-1}$, $B = 9.4 \pm 0.9$\,s$^{-1}$,
$U = 4.0 \pm 0.4$\,s$^{-1}$, $UVW1 = 0.83 \pm 0.08$\,s$^{-1}$,
$UVM2 = 0.29 \pm 0.03$\,s$^{-1}$, and $UVW2 = 0.49 \pm 0.05$\,s$^{-1}$.

Figure 1 shows the host-subtracted optical and UV light curves ASASSN-14li.\\

\noindent \textbf{Fits to the UVOT/UVM2 light curve}

The UVM2 filter provides the most robust trace of the mass accretion
rate in a TDE like ASASSN-14li; it has negligible transmission at
optical wavelengths\citemeth{Poole08,Breeveld11}.  Fits to the UVM2
light curve with a power-law of the form $f(t) = f_0 \times
(t+t_0)^{-\alpha}$ with a fixed index of $\alpha = -5/3$ imply a
disruption date of $t_0 = 56980\pm 3$ (MJD).  This model achieves a
fair characterization of the data; high fluxes between days 80-100 (in
the units of Figure 1) result in a poor statistical fit ($\chi^{2}/\nu
= 1.7$, where $\nu = 54$ degrees of freedom).  If the light curve is
fit with a variable index, a value of $-2.6\pm 0.3$ is measured (90\%
confidence).  This model achieves an improved fit ($\chi^{2}/\nu =
1.4$, for $\nu = 53$ degrees of freedom), but it does not tightly
constrain the disruption date, placing $t_0$ in the MJD 56855--56920
range.  That disruption window is adjacent to an interval wherein the
ASAS-SN monitoring did not detect the source\cite{Jose14}, making it
less plausible than the fit with $\alpha = -5/3$.

The optical bands appear to have a shallower decay curve than the UV
bands.  Recent theory\citemeth{Lodato11} predicts that optical light
produced via thermal disk emission should show a decay consistent with
$t^{-5/12}$; this might also be due to
reprocessing\citemeth{Guillochon:2014a}.  The V-band data are
consistent with this prediction, though the data are of modest quality
and a broad range of decays are permitted.\\

\noindent \textbf{X-ray monitoring observations and data reduction} 

The {\it Swift}/XRT\citemeth{Burrows05} is a charge-coupled device
(CCD).  In such cameras, photon pile-up occurs when two or more
photons land within a single detection box during a single frame time.
This causes flux distortions and spectral distortions to bright
sources.  Such distortions are effectively avoided by extracting events from
an annular region, rather than from a circle at the center of the
telescope PSF.  We therefore extracted source spectra from annuli with
an inner radius of 12 arc seconds (5 pixels), and an outer radius of
50 arc seconds.  Background flux was measured
in annular region extending from 140 -- 210 arc seconds.

Standard redistribution matrices were used; an ancillary response
file was created with the \texttt{xrtmkarf} tool utilizing a
vignetting corrected exposure map.  The source spectra were rebinned to
have 20 counts per bin with \texttt{grppha}.
In all spectral fits, we adopted a lower spectral bound of 0.3~keV
(36~$\AA$).  The upper bound on spectral fits varied
depending on the boundary of the last bin with at least 20
counts; this was generally around 1~keV (12~$\AA$).

The XRT spectra were fit with a model consisting of absorption in the
Milky Way of a blackbody emitted at the redshift of the TDE, i.e.,
\texttt{pha(zashift(bbodyrad))}, where $N_H \equiv 4 \times
10^{20}~cm^{-2}$ and $z \equiv 0.0206$.  The evolution of the best-fit
temperature of this blackbody component is displayed in Figure 3.

The blackbody temperature values measured from the
\emph{Swift}/XRT are slightly higher (kT $\simeq$ 7--10~eV) than those
measured with \emph{XMM-Newton} and \emph{Chandra}.  If an outflow
component with fiducial parameters is included in the spectral model
anyway, the XRT temperatures are then in complete agreement with those
measured using \emph{XMM-Newton} and \emph{Chandra}.\\

\noindent \textbf{Estimates of the black hole mass} 

Luminosity values inferred for the band over which the high-resolution
spectra are actually fit, and a broader band are listed in Table 1.
Taking the broader values as a proxy for a true bolometric fit, the
highest implied soft X-ray luminosity is measured in the last {\emph
  XMM-Newton} monitoring observation, giving $L \simeq 3.2\times
10^{44}~ {\rm erg}~ {\rm s}^{-1}$.  The Eddington luminosity for
standard hydrogen-rich accretion is $L_{Edd} = 1.3\times 10^{38} {\rm
  erg}~ {\rm s}^{-1} (M/M_{\odot})$.  This implies a black hole mass
of $M \simeq 2.5\times 10^{6}~ M_{\odot}$.  

Blackbody continua imply size scales, and - assuming that optically
thick blackbody emission can only originate at radii larger than the
innermost stable circular orbit (ISCO) - therefore masses.  For a
non-spinning Schwarzschild black hole, $r_{ISCO} = 6~ {\rm GM/c}^{2}$.
The blackbody emission measured in fits to the time-averaged
\emph{XMM-Newton} ``long stare'' gives an emitting area of
$3.7\times 10^{25}~ {\rm cm}^{2}$; implying $r = 1.7\times 10^{12}~
     {\rm cm}$ for a spherical geometry.  The actual geometry may be
     more disk-like, but the inner flow may be a thick disk that is
     better represented by a spherical geometry.  If the black hole
     powering ASASSN-14li is not spinning, this size implies a black
     hole mass of $M \simeq 1.9\times 10^{6}~ M_{\odot}$.

We also estimated the mass of the black hole at the heart of
ASASSN-14li by fitting the host-subtracted light curves (see Figure 1)
using the Monte Carlo software {\tt TDEFit}
\citemeth{Guillochon:2014a}.  This software assumes that emission is
produced within an elliptical accretion disk where the mass accretion
rate follows the fallback rate \citemeth{Guillochon:2013a} onto the black
hole with a viscous delay \citemeth{Guillochon:2015b}.  This emission is
then partly reprocessed into the UV/optical by an optically thick
layer \citemeth{Loeb:1997a}.
Super-Eddington accretion is treated by presuming a fitted fraction of
the Eddington excess is converted into light that is reprocessed by
the same optically thick layer.  This excess can be
produced either with an unbound wind \citemeth{Strubbe:2009a,Vinko:2015a},
or with the energy deposited by shocks in the circularization process
\citemeth{Shiokawa:2015a,Piran:2015c}.

The software performs a maximum-likelihood analysis to determine the
combinations of parameters that reproduce the observed light curves.
We utilize the ASASSN, UVOT, and XRT data in our light-curve fitting;
the most-likely models produce good fits to all bands simultaneously.
Within the context of this TDE model, a black hole mass of
0.4--1.2$\times 10^{6}~ M_{\odot}$ (1$\sigma$) is dervived.\\

\noindent \textbf{Spectroscopic observations, data reduction, and analysis} 

Table 1 lists the observation identification number (ObsId), start
time, and duration of all of the \emph{XMM-Newton} and \emph{Chandra}
observations considered in our work.  

The \emph{XMM-Newton} data were reduced using the standard Science
Analysis System (SAS version 13.5.0) tools and the latest calibration
files.  The ``rgsproc'' routine was used to generate spectral files
from the source, background spectral files, and instrument response
files.  The spectra from the RGS-1 and RGS-2
units were fit jointly.  Prior to fitting models, all
\emph{XMM-Newton} spectra were binned by a factor of five for clarity
and sensitivity.

The \emph{Chandra} data were reduced using the standard Chandra
Interactive Analysis of Observations (CIAO version 4.7) suite, and the
latest associated calibration files.  Instrument response files were
constructed using the ``fullgarf'' and ``mkgrmf'' routines.  The
first-order spectra from each observation were combined using the tool
``add\_grating\_orders'', and spectra from each observation were then
added using ``add\_grating\_spectra''.

The spectra were analyzed using the ``SPEX'' suite version
2.06\cite{Kaastra96}.  The fitting procedure minimized a $\chi^{2}$
statistic.  The spectra are most sensitive in the 18--35~\AA~ band,
and all fits were restricted to this range.  Within SPEX, absorption
from the interstellar medium in the Milky Way was modeled using the
model ``hot''; a separate ``hot'' component was included to allow for
ISM absorption within PGC 043234 at its known red-shift (using the
``reds'' component in SPEX).  The photoionized outflow was modeled
using the ``pion'' component within the SPEX suite.

Pion\cite{Kaastra96} includes numerous lines from intermediate charge
states that are lacking in similar astrophysics packages.  The fits
explored in this analysis varied the gas column density ($N_{\rm H,
  TDE}$), the gas ionization parameter ($\xi$, where $\xi = L /
nr^{2}$, and $L$ is luminosity, $n$ is the hydrogen number density,
and $r$ is the distance between the ionizing source and absorbing
gas), the rms velocity of the gas ($v_{\rm rms}$), and the bulk shift
of the gas relative to the source, in the source frame ($v_{\rm
  shift}$).

Spectra from segments within the ``long stare'' made with
\emph{XMM-Newton} were made by using the SAS tool ``tabgtigen'' to
create ``good time interval'' files to isolate periods within the light
curves of the RGS data.

The \emph{Chandra}/LETG spectra were dispersed onto the HRC, which has
a relatively high instrumental background.  Fitting
the spectra only in the 18--35\AA~ band served to limit the
contributions of background.  Nevertheless, the \emph{Chandra} spectra
are less sensitive than the best \emph{XMM-Newton} spectra of
ASASSN-14li (see Figure 2).  Prior to fitting, spectra from the two
exposures were added and then binned by a factor of three.

Figure 2 includes plots of the $\Delta \chi^{2}$ goodness-of-fit
statistic as a function of wavelength, before and after including pion
to model the ionized absortpion.  There is weak evidence of emission
lines in the spectra, perhaps with a P-Cygni profile (see below).  The
best-fit models for the high-resolution spectra predict one absorption
line at 34.5\AA~ (H-like C VI) that is not observed; small variations
to abundances could resolve this disparity.
 
Blue-shifts as small as $200~ {\rm km}~ {\rm s}^{-1}$ are measured in
the \emph{XMM-Newton}/RGS using the pion model.  According to the
\emph{XMM-Newton} User's Handbook, available through the mission
website, the absolute accuracy of the first-order wavelength scale is
6~m\AA.  At 18~\AA, this corresponds to a velocity of $100~ {\rm km}~
{\rm s}^{-1}$; at 35~\AA, this corresponds to a velocity of $51~ {\rm
  km}~ {\rm s}^{-1}$.  The model predicts numerous lines
across the 18--35~\AA~ band that are clearly detected; especially with
this leverage, the small shifts we have measured with
\emph{XMM-Newton} are robust.  In particular, the difference in
blue-shift between the low and high flux phases of the long stare,
$-360\pm 50~ {\rm km}~ {\rm s}^{-1}$ versus $-130^{-50}_{+70}~ {\rm
  km}~ {\rm s}^{-1}$, is greater than absolute calibration
uncertainties.  Differences observed in the outflow velocities between
\emph{XMM-Newton} observations are as large, or larger, and also
robust.

The lower sensitivity of the \emph{Chandra} spectra is evident in the
relatively poor constraints achieved on the column density of the
ionized X-ray outflow (see Table 1).  Similarly, the relatively high
outflow velocity measured in the \emph{Chandra} spectra, should be
viewed with a degree of caution.  The outflow velocity changes from
$\simeq 500~ {\rm km}~ {\rm s}^{-1}$ to just $-130\pm 130~ {\rm km}~
{\rm s}^{-1}$, for instance, when the binning factor is increased from
three to five.  We have found no reports in the
literature of a systematic wavelength offset between contemporaneous
high-resolution spectra obtained with \emph{XMM-Newton} and \emph{Chandra}.

The small number of high-resolution spectra complicates efforts to
discern trends.  The velocity width of the absorbing gas is fairly
constant over time, but there is a general trend toward higher
blue-shifts.  There is no clear trend in column density or ionization
parameter with time.\\

\noindent \textbf{Diffuse gas mass, outflow rates, and filling factors}

There is no a priori constraint on the density of the absorbing gas.
Taking the maximum radius implied by variability within
\emph{XMM-Newton} long stare, $r \leq 3\times 10^{15}~ {\rm cm}$, and
manipulating the ionization parameter equation ($\xi = L n^{-1}
r^{-2}$, where $L$ is the luminosity, $n$ is the number density, and
$r$ is the absorbing radius), we can derive an estimate of the
density: $n \simeq 2\times 10^{9}~ {\rm cm}^{-3}$.  Even assuming a
uniformly-filled sphere out to a radius of $r = 3\times 10^{15}~ {\rm
  cm}$, a total mass of $M \simeq 4\times 10^{32}~{\rm g}$ is implied,
or approximately $0.2~ M_{\odot}$.

The true gas mass within $r$ is likely to be orders of magnitude
lower, owing to clumping and a very low volume filling factor.  Using
the measured value of $N_{\rm H,TDE}$ and assuming $n \simeq 2\times
10^{9}~ {\rm cm}^{-3}$, $N_{\rm H,TDE} = n \Delta r$ gives a value of
$\Delta r \simeq 6.5\times 10^{12}~ {\rm cm}$.  The filling factor can
be estimated via $\Delta r / r \simeq 0.002$.  The total mass enclosed
out to a distance $r$ is then reduced accordingly, down to $4\times
10^{-4}~ M_{\odot}$, assuming a uniform density within $r$.  This is a
small value, plausible either for a clumpy wind or gas within a
filament executing an elliptical orbit.

Formally, the mass outflow rate in ASASSN-14li can be adapted from the
case where the density is known, and written as:

$\dot{M}_{\rm out} = \mu m_{p} \Omega L v C_{v} \xi^{-1}$,

\noindent where $\mu$ is the mean atomic weight ($\mu = 1.23$ is
typical), $m_P$ is the mass of the proton, $\Omega$ is the covering
factor ($0 \leq \Omega \leq 4\pi$), $L$ is the ionizing luminosity,
$v$ is the outflow velocity, $C_{v}$ is the line-of-sight global
filling factor, and $\xi$ is the ionization parameter.  Using the
values obtained in fits to the \emph{XMM-Newton} long stare (see Table
1), for instance, $\dot{M}_{\rm out} \simeq 7.9\times 10^{23}~ \Omega~
C_{v}~ {\rm g}~ {\rm s}^{-1}$.  Taking the value of $C_{v}$ derived
above, an outflow rate of $\dot{M}_{\rm out} \simeq 1.5\times 10^{21}~
\Omega~ {\rm g}~ {\rm s}^{-1}$ results.  The kinetic power in the
outflow is given by $L_{kin} = 0.5 \dot{M} v^{2}$; using the same
values assumed to estimate the mass outflow rate, $L_{kin} \simeq
3.3\times 10^{35}~ {\rm erg}~ {\rm s}^{-1}$.\\

\noindent \textbf{Emission from the diffuse outflow}

We synthesized a plausible wind emission spectrum by coupling the pion
and ``hyd'' models within SPEX.  They hyd code enables spectra to be
constructed based on the output of hydrodynamical simulations.  As
inputs, it requires the electron temperature and ion concentrations
for a gas; these were taken from our fits with pion.  We included the
resulting emission component in experimental fits to the
\emph{XMM-Newton} long stare.  The best-fit model gives an
emission measure of $1.0\pm 0.3 \times 10^{64}~ {\rm cm}^{-3}$, a
red-shift (relative to the host) of $270^{+350}_{-150}~ {\rm km}~ {\rm
  s}^{-1}$, and an ionization parameter of log$(\xi) = 4.3\pm 0.1$.

Via an F-test, the emission component is only required at the
$3\sigma$ level; however, it has some compelling
properties.  Combined with the blue-shifted absorption spectrum, the
red-shifted emission gives P-Cygni profiles.  For the gas density of $n
\simeq 2\times 10^{9}~ {\rm cm}^{-3}$ derived previously, the emission
measure gives a radius of $\simeq 10^{15}~ {\rm cm}$, comparable to
the size scale inferred from absorption variability.

The strongest lines predicted by the emission model include He-like O
VII, and H-like charge states of C, N, and O.  This model does not
account for other emission line-like features in the spectra, which
are more likely to be artifacts from spectral binning, or calibration
or modeling errors.  Emission features in the O K-edge region may be
real, but caution is warranted.  Other features are more easily
discounted as they differ between the RGS-1 and RGS-2 spectra.\\

\noindent \textbf{Code Availability}

All of the data reduction and spectroscopic fitting routines and
packages used in this work are publicly available.

The light curve modeling package, {\tt TDEFit}
\citemeth{Guillochon:2014a}, is proprietary at this time owing to
ongoing code development; a public release is planned within the
coming year.

\bibliographymeth{meth}

\end{document}